# The string
# string theorists forgot to notice
(La corde que les physiciens des cordes ont oublié de voir)

## Jean-Paul Auffray
Ex CIMS-NYU
jpauffray@yahoo.fr
(November 2007)

**Abstract**. A conjecture formulated by Henri Poincaré in 1912 suggests that Max Planck's elementary quantum of action constitutes an authentic "atom of motion". When the conjecture is implemented, the resulting entity, the *xon,* becomes "the string that string theorists forgot to notice". If a string exists in Nature, then string theoretical research should be pursued.

## 1. INTRODUCTION

Three major books concerning string theories have recently been published [1, 2, 3]. They share a characteristic: the word "action" is missing in their extensive Index. This is somewhat disconcerting since a proper understanding of action is essential to the formulation of any string theory. In this paper I resurrect a conjecture concerning action that French mathematician Henri Poincaré formulated shortly before his untimely passing in 1912. I then examine the logical consequences that ensue if one assumes the conjecture to be true and I show that, when taken at face value, the conjecture leads to the discovery of the existence of a string in Nature.

## 2. HENRI POINCARÉ'S 1912 CONJECTURE CONCERNING ACTION

Leibniz's invention of his *Dynamica* in 1689, at the heart of which he placed the abstract concept of action (*actio*), provided an alternate framework to Newton's System of the World in which Absolute Space and Absolute Time enter as primary or "God-given" principles. The subsequent discovery of the principle of least action, followed by the invention of the elementary quantum of action by Max Planck in 1900, has placed action at the very heart of modern physics.

As a physical entity worth of consideration *per se,* the elementary quantum of action was investigated by Henri Poincaré in 1912 when he proposed that it constitutes a "*véritable atome*" – an "atom of motion" – whose integrity arises from the fact that the "points" it contains are equivalent to one another from the standpoint of probability [4]. I build on this premise in what follows.

## 3. ORIGIN OF THE CONJECTURE

Poincaré's action conjecture has its roots in a remark originally made by Max Planck concerning his discovery of a connexion between the quantum of action and the Liouville Theorem of classical physics reformulated in the framework of Gibbs' statistical mechanical method of representations in phase space. Poincaré apparently first heard of the connexion when he met Planck at the Conseil de Physique ("First Solvay Congress") gathered in Bruxelles in October-November 1911. Transcripts of the Conseil meetings show that Poincaré was inquisitive of the true nature of Planck's action element all through that week [5]. The issue divided the Conseil into three "camps": those, a majority, who did not care one way or the other; those, a minority led by Arnold Sommerfeld, who considered Planck's action element to be the true fundamental entity one needed to take into account in the formulation of emerging quantum theories; and those, led by Einstein, who favoured the "energy quantum" over the action element. Throughout the Conseil meetings, Sommerfeld defended heartily the concept of action in general and that of the action element in particular, envisioning a quantum description of atomic and molecular processes in which exchanges of action elements – not of energy quanta – would be the determining factor. Opposite to him, Einstein, who spoke last, ignoring action in general and the elementary quantum of action in particular, referred to $h$ as "the second of Planck's two universal constants" and stressed the predominance of $hv$ over $h$. Thus began a programmed decline of the action element as an entity endowed with a deep physical significance of its own and the rise in its place of the more innocuous "Planck's constant", $h$.

Poincaré was obviously impressed by Planck's connexion of the action element with the Liouville Theorem and by Sommerfeld vigorous defence of the deep physical significance he attached to it. He formed his conjecture shortly after the Conseil ended. To clarify the issue involved, I shall briefly outline how the connexion arises.

# 4. CONNEXION WITH THE LIOUVILLE THEOREM

Consider two allowed states of a given (classical) physical system. If one of the states is a necessary consequence of the other, then the two states are equally probable. Let $q$ represent a generalized coordinate of the system and $p$ the corresponding conjugate momentum, then $dpdq$ constitutes an infinitesimally small elementary domain of probability associated with the system. Planck's quantum hypothesis consists in assuming that, rather than being infinitesimally small, the elementary domains are all equal and given by the relation $\iint dpdq = h$, where $h$ is Planck's famous "second universal constant" (the first is $k$, which has the dimension of entropy and intervenes to make $k$T, where T is the system temperature, an energy). Poincaré's reasoning concerning the structure of Planck's finite domains is striking: "These domains are indivisible, he wrote. If we know the system to be in one of them, then everything is automatically determined. If it were not, if events that are to follow were not fully determined by that knowledge – in a word if they were to differ depending on the system being in such or such part of the domain –, then, since the probability of some future events would not be the same in its diverse parts, the domain considered would not be indivisible from the viewpoint of probability." And he concluded: "This means that all the system states that correspond to a given domain are undistinguishable from one another and therefore constitute one and the same state." A conclusion that led him to assert unambiguously: "We are therefore led to formulate the following fundamental theorem: *A physical system can exist only in a finite number of possible distinct states; it jumps from one of these states to another without passing through intermediate states.*"[6] Thus was born in 1912, under Poincaré's pen, the concept of quantum jumps.

What interests me for the purpose of the present paper is Poincaré's suggestion that the finite domains referred to above constitute an assembly of "points".

# 5. *i*-POINTS

*We need a theory about  
what makes up space.  
Lee Smolin* [7]

Classical quantum theories describe elementary particles propagating and interacting in spacetime. Thus, even though quantum physicists do not generally acknowledge it explicitly, spacetime, made up of "points", enters the theory as a fundamental postulate. String theorists prefer to speak of one-dimensional strings propagating and interacting against a "background" that is not necessarily spacetime-like but is also made up of some kind of "points". In *The Trouble with Physics*, Lee Smolin writes: "Many quantum-gravity theorists believe there is a deeper level of reality, where space does not exist (this is taking background independence to its logical extreme)" and he offers this advice (his italics): "*Don't start with space, or anything moving in space.* Start with something that is purely quantum-mechanical and has, instead of space, some kind of purely quantum structure." [8]

In what follows I examine the possibility that Planck's action element, understood in the sense proposed by Poincaré, should be regarded to be, not simply a "universal constant", but *the* fundamental principle of quantum physics (I use the word *principle* in the Aristotelian sense it originally had: "that which comes first" [*archē*]). To implement this scheme, I shall take as a starting point Aristotle's recommendations concerning the choice of principles: "All recognize opposites as principles." To which he added: "This is indeed as it should be, for principles should not derive from one another, nor derive from other things. One should demand, to the contrary, that all the rest should derive from principles. These are precisely the conditions that primitive opposites satisfy. Being opposites, they do not derive from one another." [9]

In accordance with these precepts, I postulate the existence in nature of a subjacent physical reality arising from two opposite primitive principles:

1) An unstructured chaotic or "passive" *substrate* composed of dimensionless elements I shall designate as "$i$-points" pending further examination of their nature.
2) A primitive or "active" principle that I shall designate as $h$ since, as we shall see, it corresponds to Planck's 1900 discovery of the elementary quantum of action.

To proceed further, I shall make use of two fundamental axioms proposed by Leibniz in his magisterial treatise *Dynamica de Potentia*:

Axiom 1. "One can understand what action in a body is only in an indirect way"

Axiom 2. "Motion's formal actions are the results of the composition of diffusions and intensions." [10]

To the two axioms, I shall add a third:

Axiom 3. The interaction of $h$ with the chaotic substrate – that I shall call the "occurrence" of $h$ in the substrate – generates, or at least can generate in the substrate primitive forms of self-organisation that I shall temporarily call $i$-points *clusters*.

# 6. STANDARD REPRESENTATIONS OF *h*

In accordance with Leibniz's Axiom 2, I shall describe the occurrence of *h* in the chaotic substrate as resulting from the "composition" or "product" of two factors, one *extensive*, the other *intensive*. To give substance to this scheme, I shall invoke known quantum theoretical relations involving *h*.

I assume, first, that the occurrence of *h* in the substrate can induce in the chaotic substrate a characteristics extension, or "length", in accordance with Louis de Broglie's 1923 discovery of the relation $p = h/\lambda$ that associates a wavelength $\lambda$ to the electron momentum *p*.

In Louis de Broglie's original formulation this wavelength was regarded as a characteristic of the electron motion. I shall interpret it instead as describing a fundamental property of the action element regarded as an entity in its own rights and, accordingly, I shall write the De Broglie relation as $lp=h$ rather than $p = h/\lambda$, as in the original formulation. I shall refrain further from recognizing *a priori* one-dimensionality to the characteristic "length" *l* thus defined and shall consequently avoid calling it a wavelength. In fact, it will appear as a one-dimensional length or wavelength only under certain conditions, as we shall see. To help visualize what is involved, I shall outline a representation of the process even though it is not strictly applicable to the proposed scheme since it involves geometrical concepts not explicitly present in the scheme.

Let ABC designate an equilateral triangle supposedly representing an *h*-induced *i*-points cluster. "Distances" between any two *i*-points in this cluster are not defined. Nevertheless, for the purpose of the demonstration, let us assume that for each *i*-point in the cluster distances are defined with respect to the three arbitrarily selected *i*-points, A, B and C, which define the cluster. Let 1 designate an *i*-point chosen randomly within the cluster. Consider the *substitution* that replaces the *i*-point 1 by the *i*-point 2 located at mid-distance between 1 and B. Let a new substitution replace the *i*-point 2 thus identified by the *i*-point 3 located at mid-distance between *i*-point 2 and *i*-point C. Each subsequent substitution advances the original *i*-point midway in the direction of A, B or C, the direction being chosen at random. Somewhat surprisingly, if pursued *ad infinitum*, this process transforms the random distribution of *i*-points in the original cluster into a *fractal* distribution of *i*-points resembling a Sierpinski triangle [11]. This construction illustrates the fact that a primitive form of self-organisation can result from a (partially) random substitution process occurring within an undifferentiated substrate.

I should like to stress that in constructing this illustrative example I selected on purpose the group-theoretical concept of "substitution" to designate the self-ordering process: being equivalent to one another, *i*-points can be substituted to one another without disrupting the cluster integrity.

## 7. QUANTUM FLUCTUATIONS
## IN STRING THEORY BACKGROUNDS

The De Broglie relation, which I wrote as $lp = h$, may also be written in the alternative form $Ed = h$, in which $d$ designates a "time-like extension" or "duration". The extension associated with the occurrence of $h$ via the De Broglie original formulation thus lends itself to a dual interpretation. I shall take advantage of this duality to examine how spacetime might arise from my premises. Indeed, one cannot assume that $h$-induced clusters are permanent features of the chaotic substrate, because, if this were allowed to happen, the chaotic substrate would loose its characteristics of being fundamentally chaotic and unstructured (deprived of geometry). When considered together, the two relations $lp = h$ and $Ed = h$ describe a chaotic substrate subjected, on the average, to what might appropriately be called *quantum fluctuations*. The difference with conventional quantum theory is that we are not dealing here with a dimensioned subjacent spacetime continuum, but with a fundamentally unstructured substrate that acquires structure (geometry) and (fractal) dimensionality only in response to the action of an embedded primitive active principle acting as a *cause*.

## 8. ESOTERIC REPRESENTATIONS OF *h*

In Quantum Electrodynamics (QED), the electron action is written as

$$S = 2\pi(lp - Ed + \varphi\sigma + e\chi),$$

a formula that, besides $lp$ and $Ed$, contains two additional representations of the action element, namely $\varphi\sigma$ and $e\chi$, where $\varphi$ measures an angle, $\sigma$ the corresponding conjugate angular momentum (an action), and where e designates the electron charge and $\chi$ a gauge function.

There is more. A fifth representation of $h$ arises from the relation $e^2/2\alpha c = h$ which connects $h$ with the electron charge e, the fine structure constant $\alpha$, a pure number, and the limit velocity c (also called velocity of light in vacuum) that occurs in Poincaré's relativistic electron dynamics [12]. In this representation, the part that constitutes Leibniz's *extensio* factor is not easily identified. In the relation $e^2/2\alpha c = h$, e can vary if the fine structure constant $\alpha$ varies accordingly. Other representations of the action element will ensue if one writes other known quantum mechanical relations in the form "something = $h$". Each new representation provides a particular insight into the complex and deep nature of the action element.

# 9. ON THE NATURE OF *i*-POINTS

Any physical theory must ultimately be related to what I shall call "reality as classical physicists see it" which constitutes a particular *representation* or *interpretation* arising from the underlying reality. For the purpose of my demonstration, I take the view that "reality as classical physicists see it" depends primarily on the assumed validity of the Euclidian geometry axiom according to which there exist in nature *straight line segments* whose *lengths* can be *measured* by means of fixed-lengths rods taken as *units*. On this basis, I seek to establish a connexion between the assumed underlying chaotic "reality" in which action is the active ingredient and ordinary spacetime physics.

There is no compelling reason to assume that h-induced extensions – *l* or *d* – relate to "Euclidian" structures. I will assume instead that they relate at best to fractal structures. Let D be the fractal dimension and let $\delta$ designate the *resolution dimension* defined as the underlying fractal dimension minus the topological (observable) dimension: $\delta = D - D_T$. The use of the word *resolution* in this context refers to the following. The relations *lp = h* and *Ed = h* allow any value of *l* or *d* to occur. When particular values prevail, I will say, following a suggestion due to Laurent Nottale, that they specify the cluster *state of resolution* [13]. This assigns to the cluster a new kind of coordinate which is neither space-like nor time-like but reflects instead a *fundamental* dependence of the cluster geometry on resolution. To specify more closely the significance of this concept, I shall reason concurrently within two frameworks: 1) the framework that arises from the 2-principle underlying scheme; and 2) the framework of ordinary four-dimensional spacetime physics in which distances between events (world-points) are the relativistic intervals that connect them. To simplify the presentation however, I shall use the word "length" rather than the word "interval" to designate distances between events (points) in spacetime.

Distances can be measured in spacetime only if one disposes of "yardsticks" to make the measurements. Let the extension *l* associated with an *h*-induced *i*-points cluster be $l(\varepsilon)$ when the yardstick length is $\varepsilon$ in spacetime. Laurent Nottale's Scale Relativity Theory (SR) assumes that $l(\varepsilon)$ tends to infinity when $\varepsilon$ tends to zero. The yardstick length $\varepsilon$ is only defined relatively to the length $\varepsilon'$ of another yardstick, however. In the simplest formulation of the theory, the resolution dependence of *l* assumes the simple form $l' = (\varepsilon/\varepsilon')^\delta . l$. When $\delta = 1$, which arises when the fractal dimension is 2 and the topological dimension is 1, the resolution dependence becomes $l'/l = \varepsilon/\varepsilon'$. Inasmuch as it corresponds to the supposition $t'/t = 1$ which characterises classical mechanics, the supposition $\delta = 1$ may be said to impart to this formulation a *Galilean* character. In a more sophisticated formulation of the theory a *transition scale* $\varepsilon_0$ occurs such that for $\varepsilon \ll \varepsilon_0$ the system is fundamentally resolution-*dependent*, while for $\varepsilon \gg \varepsilon_0$, it becomes essentially resolution-*independent*. The transition scale thus defined establishes a connexion between quantum and classical behaviour, a highly desirable feature for a scheme that seeks to define a new kind of background for string propagation and interaction.

## 10. CONTINUITY AND DIMENSIONALITY

In his article *On the Foundations of Geometry* published in 1898 (and expanded further en 1903), Henri Poincaré investigates the reason we perceive space as being a three-dimensional *continuum*. He ascribes the phenomenon not to a requirement of space itself, but to the way we combine and interpret "impressions" we receive from the external world [14]. Briefly outlined, his argument runs as follows.

Divers *impressions systems*, originated in the external world, are transmitted to our brain, which then compares them among themselves. Often, two of these systems cannot be distinguished from one another, or they can be distinguished from one another yet not from a third system intermediate to both. When this happens, these systems, taken together, generate a "physical continuum", each impression system being an *element* of the continuum. Poincaré gave a concrete example of the scheme thus established. Assume we can distinguish object A, measuring 10 mm, from object C, measuring 12 mm, but cannot distinguish either one from object B, measuring 11 mm. *For us*, A, B, C, … then constitute a continuum, even though this continuum does not necessarily possess *objective reality*. Indeed, its structure leads to intolerable logical difficulties. The example given above, for instance, yields the relations A=B, B=C, A<C. To alleviate this logical inconsistency, Poincaré introduced the notion of a *mathematical continuum* in which the constitutive elements do not overlap each other as they do in the physical continuum, leading to the logically acceptable relations A<B, B<C, A<C, in the example considered.

Applied to i-points clusters, this analysis suggests why/how they might appear *to us* as forming a continuum.

## 9. A QUESTION OF NOTATION

According to Max Planck's recollection, his fateful encounter with dynamical action occurred on Sunday, October 7, 1900 as he was taking his eldest son Karl for a walk in a nearby park. As he described later the incident, Max Plank told his son: "I just made a discovery which might equal in significance that made by Newton [universal gravitation!]." To construct his black body radiation formula Max Planck needed a pure number that he could enter as the exponent of an exponential function. He chose to construct it as the ratio of two factors having each the dimension of energy, one proportional to the radiation frequency $v$, the other proportional to the black body temperature T. To that effect, he introduced in the calculation two "universal constants", $h$ and $k$, such that the products $hv$ and $k$T would both have the dimension of energy. The ratio $hv/k$T then provided him with the pure number he needed. Reflecting later on his initiative, Planck noted that his constant $h$ had the dimension of action. To name the constant $h$ Planck selected the German word Wirkung, which literally means "effect", rather than words such as Aktie or Aktion, which might have been more appropriate. Planck also noticed that the constant $h$ measures a "fixed quantity" of "Wirkung". He therefore appended to Wirkung the germanised Latin word "quantum", which means simply "quantity", yielding the designation Wirkungsquantum. When he saw further that the newborn Wirkungsquantum had the characteristic of a fundamental "unit" or "element", he completed the designation with the qualifier Elementares, so that the final designation became Das Elementares Wirkungsquantum – for us today the elementary quantum of action, also called the action element.

[Why did Max Planck not choose to call his discovery Das Elementares *Aktions*quantum (or *Akties*quantum)? I shall venture a possible "explanation": a (brilliant) musician, Max Planck may have felt that Wirkungsquantum sounded better than Aktiequantum or Aktionquantum (does it?).] [15]

Having forged the name Elementares Wirkungsquantum to designate the discovery and selected the letter $h$ to represent it in formulae, Max Planck was left with another fundamental choice: should he represent the "energy quantum" corresponding to the radiation frequency $v$ as $hv$ or should he prefer the notation $vh$? He chose $hv$… when he should have chosen $vh$ – for the following simple reason. Assign a numerical value to $v$ – "one thousand per second" for instance. Then, translated into words, $vh$ reads: "This black body emits one thousand Elementares Wirkungsquanta per second." The inverse construction $hv$ does not properly translate into words.

Planck's choice of $hv$ over $vh$ to represent the energy quantum associated to the frequency $v$ has had a devastating impact on the development of the "quantum theories" which flourished into being soon after the introduction of $h$ into physics was accepted by the physics community.

Once Planck's universal constant $h$ is restored to the status of an action element constituting a "véritable atome" in accordance with Henri Poincaré's 1912 conjecture, the "energy quantum" $hv$ and its derivative, the photon, disappear as such from the physical theory and the proper designation $vh$ takes on its full meaning: radiation consists in the emission or absorption of action elements. Indeed, this is precisely what Arnold Sommerfeld stated in Bruxelles in 1911 when he formulated this relativistic quantum postulate: "*To each natural transformation there corresponds a number of action elements that is independent of the referential system* (i. e. is a particular multiple of *h*)."

## 11. THE STRING THAT STRING THEORISTS FORGOT TO NOTICE

If Nature's underlying reality can indeed be described in terms of a background that derives its (fractal) structures from the occurrence of an action element acting on its "points" as a "first principle" or *cause*, then significant consequences will ensue. One is that the symbol *h*, which represents the action element, should no longer be allowed to enter quantum mechanical or string theoretical equations in any odd way. In fact, all quantum mechanical and string theoretical equations in which *h* appears will have to be rewritten in the general form "*something = h*". This rewriting of quantum mechanics will be advantageous in many cases. Physicists will realize for instance that "angular momentum" is action in disguise (it has the dimension of action!). They will consequently accept more readily the fact that it is "quantized". Spin, which has also the dimension of action, will take on a new significance. And physicists will discover with some surprise and perhaps with some pleasure that Richard Feynman's QED consists in its most fundamental aspect in calculating the action *S* for an elementary physical process to occur, then divide *S* by *h* to yield the number *S/h* of action elements *h* comprised in the calculated action *S*, a most satisfying result from my point of view (it means that QED is compatible with the proposed new scheme).

Some years ago, as I began to explore the scheme alluded to in this paper, I assigned a name to the action element treated as an "atom of motion": I called it the *xon*. I also called *Step Mechanics* Richard Feynman's QED revisited in the light of the remark made above. Inasmuch as the *xon* is an entity to which a "length" and/or a "duration" of some sort can be assigned, it is indeed a "string" – a string from which, furthermore, spacetime variables arise. In *The Trouble with Physics,* Lee Smolin asks for "something that is purely quantum-mechanical and has, instead of space, some kind of purely quantum structure". I submit that the *xon* fulfils this wish and that it might thereby be proper to see in it the string that string theorists asked for "but forgot to notice".

November 2007.

# ANNEXE

# A brief essay on the history of action

## A1. ACTION IN CONTEMPORAY PHYSICS

In Section §1 of this paper I allude to the fact that the word "action" is missing in the Index of three major books recently published concerning strings. This is no accident. The word "action" is also missing in the Index of one of the most famous twentieth century didactic treatise concerning physics, the *Feynman Lectures on Physics.* As he delivered his celebrated *Lecture on the Principle of Least Action* at Caltech in 1964, Richard Feynman, who had (re)integrated action in physics in the 1960s, proposed this "definition" of action: "We have a certain quantity which is called *action*, S. It is the kinetic energy, minus the potential energy, integrated over time." [*The Feynman Lectures on Physics,* II, 19-3]. This presentation describes the way action is classically calculated ever since Lagrange established the formula in the eighteenth century, but it fails to provide an insight into the deep nature of the concept which lies behind the word. Incidentally, early in his *Lectures* Richard Feynman asks this question: "In the first place, what do we mean by *time* and *space*?" And he supplies this "answer": "It turns out that these deep philosophical questions have to be analyzed very carefully in physics, and this is not so easy to do." [*Lectures*, I, 8-2]

Action as a concept is at the heart of the considerations developed in this paper, not the method used for calculating it. For the benefit of those who might not be thoroughly familiar with the history of action, I present here a brief review of the way it has entered the physicist's vocabulary.

## A2. ON G. W. LEIBNIZ'S INVENTION OF DYNAMICA

In 1685, King Louis XIV of France revoked the so-called "Edit de Nantes" that his grandfather, King Henri IV, had signed in 1598 granting French Protestants the "perpetual and inalienable" right to practice their cult in the kingdom as they wished. The revocation and Louis XIV's exacerbated territorial ambitions brought Europe to the brink of war. The war broke out in 1689, involving France, Germany, Spain and England. Appalled and determined to prevent a European disaster, Gottfried Wilhelm Leibniz, then historiographer for the Duke of Brunswick, undertook a scholarly and diplomatic trip which took him, over a period of time, all the way to Rome, where he met with Jesuit scholars and with the Pope. Returning from this trip in October 1689, Leibniz stopped in Florence, where, half a century earlier, Galileo had announced to the world his formulation of "two new sciences". Inspired by Galileo's example, Leibniz formalized his own "new science", an endeavor he had thought about for some time since 1676 and that he had perfected while he traveled to Rome. He forged for his new science a name built from Greek roots, which he adapted to Latin taste: *Dynamica*. At the heart of Dynamica, he placed a new concept for which he coined the (Latin) name *actio* – "action" for us today. Thus was born one of the most mysterious conceptual tools ever imagined to help unravel the workings of Nature.

## A3. ON G. W. LEIBNIZ'S INVENTION OF THE CONCEPT OF ACTION

In inventing the concept of action Leibniz did not have in mind any "Principle of Least Action". This came later at the hands of others. On July 5, 1687, while Leibniz was on his way to Rome, Newton's 510-pages treatise *Philosophiae Naturalis Principia Mathematica* appeared in print. As he read it, Leibniz must have been satisfied to note that Newton's work lacked the depth and insights his own "new science" possessed. Rather than imitate Newton and publish a comprehensive treatise on Dynamica, Leibniz used several different means to "diffuse" his invention throughout the learned community. The diffusion met, over time, with various degrees of success. Newton rejected Dynamica outright. Others accepted readily Leibniz's new vocabulary as they accepted – Newton excepted – his notation for his invention of the Differential and Integral Calculus. Leibniz's writings concerning action are collected in volume 6 of his Mathematische Schriften [*see* ref. 10].

## A4. MAUPERTUIS DISCOVERS THE PRINCIPLE OF LEAST ACTION

I shall now consider action's resurgence, half a century later, at the hands of French philosopher Pierre Louis Moreau de Maupertuis. Called to Berlin by King Frederic II to preside over the Prussian Academy, Maupertuis bid farewell to his colleagues of the French Académie des sciences during a meeting held on April 15, 1744 in the Louvre palace. At this meeting, Maupertuis announced his discovery of a new fundamental working mechanism of nature: "When light travels from one medium to another […] the path it follows is that for which the Quantity of Action is the least." Shortly after Maupertuis arrived in Berlin, violent verbal attacks directed by Voltaire against him caused Frederic the Great to order Voltaire out of Prussia. Present in Berlin during the quarrel, Leonhardt Euler defended Maupertuis heartily and showed how to calculate action by means of an integral taken over the path of the moving body.

## A5. LAGRANGE INCORPORATES ACTION IN HIS ANALYTICAL MECHANICS

On the eve of the French Revolution in 1789, half a century after Maupertuis, Joseph Lagrange, who years before had succeeded Euler in Berlin, published his celebrated *Mechanique analitique* in which he devoted a chapter to the description of the "Principles of Dynamics", one of which, said Lagrange, is the Principle of Least Action. But Lagrange, oddly, asserted that the "Principles" he described should be regarded as expressing *consequences* of the Laws of Dynamics rather than constitute "primitive principles" of that Science. Thus, to him, action enters in a given problem of mechanics only as the *result* of an integration taken over time and does carry any significance of its own. It would take another century before action *per se* entered for good in the vocabulary of physical science. This grand entrance constitutes a story in itself, some aspects of which I have alluded to in this paper.